# Approximate energy states and thermal properties of a particle with position-dependent mass in external magnetic fields

M. Eshghi*,[1], H. Mehraban[2], S. M. Ikhdair[3,4]

[1] *Young Researchers and Elite club, Central Tehran Branch, Islamic Azad University, Tehran, Iran*

[2] *Faculty of Physics, Semnan University, Semnan, Iran*

[3]*Department of Physics, Faculty of Science, An-Najah National University, Nablus, West Bank, Palestine*

[4] *Department of Electrical Engineering, Near East University, Nicosia, Northern Cyprus, Mersin 10, Turkey*

## Abstract

We solve the Schrödinger equation with a position-dependent mass (PDM) charged particle interacted via the superposition of the Morse and Coulomb potentials and exposed to external magnetic and Aharonov-Bohm (AB) flux fields. The non-relativistic bound state energies together with their wave functions are calculated for two spatially-dependent mass distribution functions. We also study the thermal quantities of such a system. Further, the canonical formalism is used to compute various thermodynamic variables for second choosing mass by using the Gibbs formalism. We give plots for energy as a function of various physical parameters. The behavior of the internal energy, specific heat and entropy as functions of temperature and mass density parameter in the inverse-square mass case for different values of magnetic field are shown.

## 1. Introduction

The calculation of the fundamental physical quantities in many physical sciences is the first work one needs to perform. As a result, the exact solutions of the Schrödinger and Dirac wave equations have become the essential part from the beginning of quantum mechanics [1] and such solutions are also useful in the various fields of the atomic, nuclear and high energy physics [2-9].

---

* ***Email corresponding author:*** *eshgi54@gmail.com; m.eshghi@semnan.ac.ir*

In fact, the solution of the Schrödinger equation for a particle interacted via potential fields are mainly used to describe many systems of fundamental importance in the applications of quantum physics, for example, the hydrogen atom and the harmonic oscillator. Further, the exact solutions of this equation, expressed in analytical form, describing one-electron atoms are fundamental in studying the atomic structure theory. For example, in studying exactly solvable models, we can see that the analytical form of a wave function describing electron correlation of the nonrelativistic equation for two interacting electrons confined in a potential field [10]. Also, the non-relativistic equation can be reduced to the form of biconfluent Heun equation known in mathematics since a century for harmonium [11, 12]. Further, the exact analytical solutions are essentially used into quantum-chemical and quantum electrodynamics and theory of molecular vibrations. These analytical solutions of the nonrelativistic equation are also very important to examine the correctness of models and approximations in computational physics and chemistry as well. It is interesting that nearly all analytic solutions of the non-relativistic equation have been expressed in terms of hypergeometric functions [13-15]. However, for a large number of potentials of physical importance used into the Schrödinger equation or the perturbations into Dirac and Dirac-Weyl equations may be transformed into the form of the Heun equation [16-20]. On the other hand, the study of the non-relativistic equation for a particle with PDM system has been a matter of interest since the early days of solid state physics. The idea of PDM arises due to the effect of the periodic field on the non-relativistic motion of electrons periodic lattices. In fact, it happens in typical semiconductors by the effect of impurities in periodic lattices perturbed [21]. Recently, a considerable interest in the mass dependence on the inter-nuclear distance has been revived in solving the Schrödinger, Klein-Gordon, Dirac and Duffin-Kemmer-Petiau wave equations with various central potential models [22-38].

Recently, a number of works take the effects of an electric or magnetic fields into account in studying different systems [17-20, 39, 40]. In this regard, Yuce has solved analytically the Schrödinger equation for a charged particle interacting with the plane-wave electromagnetic field [41]. Unal *et al* have investigated electric field effects on the refractive index changes in a modified Poschl-Teller quantum well [42].

However, to the best of our knowledge, we report the solution of the Schrödinger equation for a particle with a spatially dependent mass in the potential field under the influence of external magnetic and Aharonov-Bohm (AB) flux fields. In fact, we find that in obtaining these solutions for the Schrödinger equation with these conditions would be very useful for some of the physical systems.

In this work, we intend to solve the Schrödinger equation with the superposition of Morse-plus-Coulomb potential taking the general form:

$$V(\rho) = V_1 exp(-\lambda\rho) + V_2 exp(-2\lambda\rho) + \frac{V_3}{\rho}, \tag{1}$$

where $V_1$, $V_2$, $V_3$ and $\lambda$ are potential constants with two physically presumed PDM distribution functions of the exponential and inverse-square forms:

$$M(\rho) = m_0 exp(-a\rho), \tag{2a}$$

and

$$M(\rho) = \frac{a}{\rho^2}, \tag{2b}$$

where $a$ is mass constant and placed in an external perpendicular magnetic and AB flux fields.

The organization of this paper is as follows. In section 2, we write the main obtained equation with a suitable change of the spatially dependent mass variable. Then we calculate the energy states and their corresponding wave functions by using the series method. At the end, we discuss the limiting case for the stationary mass and compare our analytical results with those ones obtained by other authors wherever possible. Finally, in section 3, we give our discussions and conclusions.

## 2. Solution of Schrodinger Equation for Two Physical PDM Functions

In this section, we seek to solve the Schrödinger equation for a particle with a position-dependent mass (PDM) placed in Morse-plus-Coulomb interactions potential field and under the influence of external perpendicular magnetic and AB flux fields in the cylindrical two-dimensional space. We calculate the bound state energies and their corresponding wave functions and find their thermodynamic properties. The general form of the Schrodinger equation for a particle with PDM system under the action of potential field and in the presence of the vector potential is given by

$$\left(\hat{p} + \frac{e}{c}\vec{A}\right) \cdot \frac{1}{2M(\rho)}\left(\hat{p} + \frac{e}{c}\vec{A}\right)\Psi(\rho,\varphi,z) = [E_n - V(\rho)]\Psi(\rho,\varphi,z), \tag{3}$$

where $M(\rho)$ is the effective electronic radial mass distribution and $\vec{A}$ is the vector potential which can be found in terms of the magnetic field.

Let us assume that the vector potential has the simple form: $\vec{A} = (0, B\rho/2 + \Phi_{AB}/2\pi\rho, 0)$ where $B$ and $\Phi_{AB}$ are the magnetic and AB flux fields. It has been used in quantum dot and quantum pseudot [20, 37, 43].

After a lengthy but straightforward algebra we arrive at the following second-order differential equation in the radial form:

$$\frac{d^2R(\rho)}{d\rho^2} + \left[\frac{1}{\rho} - \frac{M'(\rho)}{M(\rho)}\right]\frac{dR(\rho)}{d\rho} + \left[-\frac{eB}{\hbar c}\delta - \frac{\delta^2}{\rho^2} - \left(\frac{eB}{2\hbar c}\right)^2 \rho^2 + \frac{2M(\rho)}{\hbar^2}[E_{nm} - V(\rho)]\right] R(\rho) = 0, \quad (4)$$

where $m$ is the magnetic quantum number and $\delta = m + e\Phi_{AB}/c\hbar$ is the so called new magnetic quantum number.

## 2.1. Exponential Mass Distribution

Now we need to solve Eq. (4) by substituting Eqs. (1) and (2a) and making expansion for the exponential function up to first degree in $\rho$: $exp(-\lambda\rho) \cong 1 - \lambda\rho$, then the above differential equation (4) becomes

$$\frac{d^2R(\rho)}{d\rho^2} + \left(\frac{1}{\rho} + a\right)\frac{dR(\rho)}{d\rho} + \left[\eta_1 + \eta_2\frac{1}{\rho^2} + \eta_3\frac{1}{\rho} + \eta_4\rho + \eta_5\rho^2\right]R(\rho) = 0. \quad (5)$$

where we have used the following assignments

$$\eta_1 = -\frac{eB}{\hbar c}\delta + \frac{2m_0}{\hbar^2}(E_{nm} - V_1 - V_2 - aV_3),$$

$$\eta_2 = -\delta^2, \qquad \eta_3 = -\frac{2m_0 V_3}{\hbar^2},$$

$$\eta_4 = \frac{2m_0}{\hbar^2}\left(-aE_{nm} + (\lambda + a)V_1 + (2\lambda + a)V_2 + \frac{a^2}{2}V_3\right), \quad (6)$$

$$\eta_5 = -\left(\frac{eB}{2\hbar c}\right)^2 + \frac{2m_0}{\hbar^2}\left(\frac{a^2}{2}E_{nm} - a\lambda V_1 - 2a\lambda V_2\right).$$

Equation (5) is the standard form of a second order linear differential equation. In order to solve Eq. (5), we can transform it into a rather simpler form by using the following physical ansatz to the trial wave function:

$$R(\rho) = \frac{1}{\sqrt{\rho}} \exp\left(-\frac{a\rho}{2}\right) f_{nm}(\rho), \tag{7}$$

where the function $f(\rho)$ is found to satisfy the following differential equation:

$$\frac{d^2 f(\rho)}{d\rho^2} - \left[\eta_1 - \frac{a^2}{4} + \frac{\eta_2 + \frac{1}{4}}{\rho^2} + \frac{\eta_3 - \frac{a}{4}}{\rho} + \eta_4 \rho + \eta_5 \rho^2\right] f(\rho) = 0. \tag{8}$$

Equation (8) can be further simplified by taking the following definitions as

$$\xi = \eta_1 - \frac{a^2}{4}, \qquad -\alpha^2 = \eta_2,$$

$$c_1 = \eta_3 - \frac{a}{4}, \qquad -b_1 = \eta_4, \qquad -\varepsilon = \eta_5, \tag{9}$$

then Eq. (8) turns out to be

$$\frac{d^2 f(\rho)}{d\rho^2} + \left(\xi - \frac{\alpha^2 - 1/4}{\rho^2} + \frac{c_1}{\rho} - b_1 \rho - \varepsilon \rho^2\right) f(\rho) = 0. \tag{10}$$

Making the following change of variable: $\chi = \varepsilon^{1/4} \rho$, then the so-called the radial biconfluent Henu's equation (10) reduces into Schrödinger type equation

$$\left(\frac{d^2}{d\chi^2} + \frac{\xi}{\varepsilon^{1/2}} - \frac{\alpha^2 - 1/4}{\chi^2} + \frac{c_1}{\varepsilon^{1/4}} \frac{1}{\chi} - \frac{b_1}{\varepsilon^{3/4}} \chi - \chi^2\right) f(\chi) = 0. \tag{11}$$

The differential Equation (11) has two singular points: an irregular singular point at infinity and a regular one at the origin. In order to calculate the energy spectrum of Eq. (11), it is convenient to analyze its asymptotic behavior. Unfortunately, in the general case, there is no closed expression for large values of argument for the asymptotic behavior of the biconfluent Heun function. In order to calculate a polynomial form with the conditions on parameters of the biconfluent Heun, it can only be obtained by analyzing each coefficient separately.

Now, the behavior of the solution of Eq. (11) for $\chi \to 0$, that is, at $f(0) = 0$, can be determined by centrifugal term and the asymptotic behavior of the solution of Eq. (11) for $\chi \to \infty$, that is, at $f(\chi \to \infty) \to 0$, can be determined by the oscillator terms. In this regard, note that Eq. (11) has been analyzed for $b_1 = \varepsilon = 0$ into Ref. [43].

Now, after the above analysis, we can choose a suitable ansatz for the function $f(\chi)$ as follows:

$$f(\chi) = \chi^{\alpha + 1/2} \exp\left(-\frac{\chi^2 + \chi \tilde{b}}{2}\right) F(\chi), \tag{12}$$

then Eq. (11) can be written as

$$\chi\frac{d^2F(\chi)}{d\chi^2}+\left[2\left(\alpha+\frac{1}{2}\right)+\tilde{b}\chi-2\chi^2\right]\frac{dF(\chi)}{d\chi}$$
$$+\left\{\eta-\tilde{b}\left(\alpha+\frac{1}{2}\right)+\left[\frac{\zeta}{\gamma}+\frac{\tilde{b}^2}{4}-2\left(\alpha+\frac{1}{2}\right)-1\right]\chi\right\}F(\chi)=0, \tag{13}$$

where we have defined $\eta=c_1/\varepsilon^{1/4}$ and $\tilde{b}=b_1/\varepsilon^{3/4}$.

Note that Eq. (13) resembles the so-called biconfluent Heun's (BCH) differential equation [16, 44],

$$\frac{d^2u}{d\zeta^2}+\frac{1}{\xi}(\alpha+1-\beta\xi-2\xi^2)\frac{du}{d\xi}+\left[\frac{\delta+\beta+\alpha\beta}{2}+(\gamma-\alpha-2)\xi\right]\frac{u}{\xi}=0, \tag{14}$$

with the Heun's wave-function solution given by $u=H_B(\alpha,\beta,\gamma,\delta,-\xi)$.

At first, we introduce the parameters $\Delta_1$, $\Delta_2$, $\Delta_3$ and the function $\tilde{F}$ representing $F$ as follows:

$$\Delta_1=\left(\alpha+\frac{1}{2}\right),\qquad \Delta_2=1-\left(\alpha+\frac{1}{2}\right)\frac{\tilde{b}}{\eta},$$
$$\Delta_3=\frac{\xi}{\varepsilon^{1/2}}+\frac{\tilde{b}^2}{4}-2\left(\alpha+\frac{1}{2}\right)-1,\qquad if\ \tilde{F}=F. \tag{15}$$

Thus, Eq. (13) can be rewritten as

$$\chi\frac{d^2\tilde{F}(\chi)}{d\chi^2}+[2\Delta_1-\tilde{b}\chi-2\chi^2]\frac{d\tilde{F}(\chi)}{d\chi}+(\Delta_3\chi-\Delta_2\eta)\tilde{F}(\chi)=0. \tag{16}$$

Now, by choosing $\tilde{F}(\chi)=\sum_n A_n\chi^n$, we can solve Eq. (16) by the Frobenius series. After substituting value of $\tilde{F}(\chi)$ into Eq. (16), the following recurrence relation can be obtained as

$$A_{n+2}=\frac{[\Delta_2\eta-\tilde{b}(n+1)]A_{n+1}-(\Delta_3-2n)A_n}{(n+2)(n+2\Delta_1+1)}. \tag{17}$$

Assuming $A_{-1}=0$ and $A_0=0$, we can calculate the first three coefficients of the above recurrence relation, Eq. (17), as follows:

$$A_1=\frac{\Delta_2\eta}{2\Delta_1}A_0,$$
$$A_2=\frac{1}{2(2\Delta_1+1)}\left[\frac{\Delta_2\eta}{2\Delta_1}(\Delta_2\eta-\tilde{b})-\Delta_3\right], \tag{18}$$
$$A_3=\frac{-1}{6(\Delta_1+1)}\left\{\frac{\Delta_2\eta-2\tilde{b}}{2(2\Delta_1+1)}\left[\Delta_3-\frac{\Delta_2\eta}{2\Delta_1}(\Delta_2\eta-\tilde{b})\right]+\frac{\Delta_2\eta}{2\Delta_1}(\Delta_3-2)\right\}.$$

At this stage, we can calculate the analytical solution to the radial part of the Schrödinger equation. This work can be achieved by breaking the series (17) of the BCH function into Heun's polynomial of degree $n$. Imposing the following conditions on the two coefficients as $A_{n+1} = 0$ and $\Delta_3 = 2n$ with $n = 1, 2, 3, \cdots$ must be simultaneously satisfied. From the second condition $\Delta_3 = 2n$, it is possible to calculate formal expression for the energy eigenvalues. Therefore, after adopting this limitation $\Delta_3 = 2n$ we can simply calculate the energy eigenvalues through the energy equation

$$4\xi + \varepsilon^{1/2}(b_1^2 - 4) - 8\varepsilon^{1/2}(\alpha + 1/2 - n) = 0. \tag{19}$$

by using the Eqs. (9) and Eq. (19), the energy eigenvalue equation becomes

$$4\left[-\frac{eB}{\hbar c}\left(m + \frac{e\Phi_{AB}}{c\hbar}\right) + \frac{2m_0}{\hbar^2}(E_{nm} - V_1 - V_2 - aV_3) - \frac{a^2}{4}\right] +$$

$$\sqrt{\left(\frac{eB}{2\hbar c}\right)^2 - \frac{2m_0}{\hbar^2}\left(\frac{a^2}{2}E_{nm} - a\lambda V_1 - 2a\lambda V_2\right)} \times \left\{\left[\frac{2m_0}{\hbar^2}\left(-aE_{nm} + (\lambda + a)V_1 + (2\lambda + a)V_2 + \frac{a^2}{2}V_3\right)\right]^2 - 4 - 8\left(m + \frac{e\Phi_{AB}}{c\hbar} + \frac{1}{2} - n\right)\right\} = 0, \tag{20a}$$

where $n = 1, 2, 3, \cdots$.

Here to examine the behavior of the energy eigenvalues in Eq. (20a), we plot the energy as a function of the potential parameter, mass parameter and magnetic field as shown in Figures 1 to 4. In Fig. 1, the energy as a function of mass density parameter $a$ is plotted for various values of magnetic field strength, $B$ when $\Phi = 1.0 \, Tesla$. The energy increases exponentially with the increasing of $a$. The particle is strongly bound when $a$ is small but is less bound when $a$ is large. The magnetic field change has a slight effect on energy curve. In fact, notice that for a fixed value of $a$, the energy decreases when the magnetic field grows. Namely, increasing magnetic field makes a particle strongly bound. The effect of the magnetic flux density, $\Phi$, when $B = 1.0 \, Tesla$ on energy is stronger as they become more bound and is very similar, but greater than that of the influence of magnetic field. In fact, notice that for a fixed value of $a$, the energy decreases when the magnetic flux density grows. In Fig. 2, the behavior of energy with potential parameter $\lambda$ is linear for different values of $B$ and $\Phi$, respectively. In here, we see that the energy decrease with increasing values of $B$ and $\Phi$. In Fig. 3 and Fig. 4, the energy increases linearly with increasing magnetic field B and magnetic flux density $\Phi$, respectively, for various values of $a$. The particle becomes less positive when $a$ is increasing under the influence of both B and $\Phi$.

On the other hand, to find the energy eigenvalue equation for the stationary mass case, we put $a = 0$ in (20a) and get

$$4\left[-\frac{eB}{\hbar c}\left(m + \frac{e\Phi_{AB}}{c\hbar}\right) + \frac{2m_0}{\hbar^2}(E_{nm} - V_1 - V_2)\right] + \frac{eB}{2\hbar c}\left\{\left[\frac{2m_0\lambda}{\hbar^2}(V_1 + 2V_2)\right]^2 - 4 - \right.$$
$$\left. 8\left(m + \frac{e\Phi_{AB}}{c\hbar} + \frac{1}{2} - n\right)\right\} = 0, \tag{20b}$$

Finally, to obtain of the wave function, in comparing Eq. (13) with its counterpart Eq. (14) and using Eq. (12), we can deduce that Eq. (13) is simply the BCH differential equation [16, 44], whose solution is BCH function, $H_B$:

$$f_{nm}(\chi) = \chi^{m + \frac{e\Phi_{AB}}{\hbar c} + 1/2} \exp\left[-\frac{\chi\left(\chi + \frac{2m_0\lambda}{\hbar^2}\left(-E_{nm} + 2V_1 + 3V_2 + \frac{\lambda V_3}{2}\right)\right)}{2}\right]$$

$$\times H_B\left(m + \frac{e\Phi_{AB}}{\hbar c}, \frac{\frac{2m_0\lambda}{\hbar^2}\left(-E_{nm} + 2V_1 + 3V_2 + \frac{\lambda V_3}{2}\right)}{\varepsilon^{3/4}}, -\frac{eB}{\hbar c}\delta\right.$$

$$+ \frac{2m_0}{\hbar^2}(E_{nm} - V_1 - V_2 + \lambda V_3) - \frac{\lambda^2}{4}$$

$$\left. + \frac{\left[\frac{2m_0\lambda}{\hbar^2}\left(-E_{nm} + 2V_1 + 3V_2 + \frac{\lambda V_3}{2}\right)\right]^2}{4\varepsilon^{3/2}}, \frac{-\frac{4m_0 V_3}{\hbar^2} - \lambda}{\varepsilon^{1/4}}, -\chi\right). \tag{21}$$

At the end, using the Eq. (7) and $\chi = \varepsilon^{1/4}\rho$, we can obtain the radial wave function as

$$R(\rho) = \frac{1}{\sqrt{\rho}}\exp\left(-\frac{a\rho}{2}\right)f_{nm}(\rho). \tag{22}$$

## 2.2. Inverse-Square Mass Distribution

Here we choose the inverse-square mass distribution function [45]

$$M(\rho) = \frac{a}{\rho^2}, \tag{23}$$

where $a$ is an arbitrary non-zero real constant parameter.

After substituting both Eq. (1) (with the assumption that $V_3 = 0$) and Eq. (23) into Eq. (4) and making expansion for the exponential function to fourth degree in $\rho$: $\exp(-\lambda\rho) \cong 1 - \lambda\rho + \lambda^2\rho^2/2 - \lambda^3\rho^3/6 + \lambda^4\rho^4/24$, then the differential equation (4) becomes

$$\frac{d^2R(\rho)}{d\rho^2}+\frac{3}{\rho}\frac{dR(\rho)}{d\rho}+\left[\eta_1+\eta_2\frac{1}{\rho^2}+\eta_3\frac{1}{\rho}+\eta_4\rho+\eta_5\rho^2\right]R(\rho)=0, \qquad (24)$$

with the following identifications

$$\eta_1 = -\frac{eB}{\hbar c}\delta + \frac{\lambda^2 a}{\hbar^2}(V_1+4V_2),$$

$$\eta_2 = -\delta^2 + \frac{2a}{\hbar^2}(E_{nm}+2V_1+2V_2),$$

$$\eta_3 = -\frac{2\lambda a}{\hbar^2}(V_1+2V_2), \qquad (25)$$

$$\eta_4 = -\frac{a\lambda^3}{3\hbar^2}(V_1+8V_2),$$

$$\eta_5 = -\left(\frac{eB}{2\hbar c}\right)^2 - \frac{a\lambda^4}{12\hbar^2}(V_1+16V_2),$$

have been used. Now, setting the radial wave function as $R(\rho) = \rho^{-\frac{3}{2}}f(\rho)$, where $f(\rho)$ is obtained from the second order deferential equation as

$$\frac{d^2f(\rho)}{d\rho^2}+\left[\eta_1+\left(\eta_2-\frac{3}{4}\right)\frac{1}{\rho^2}+\eta_3\frac{1}{\rho}+\eta_4\rho+\eta_5\rho^2\right]f(\rho)=0. \qquad (26)$$

Setting the following assignments into Eq. (26) as

$$\xi = \eta_1, \qquad -\alpha^2 = \eta_2 - \frac{3}{4},$$
$$c_1 = \eta_3, \qquad -b_1 = \eta_4, \qquad -\varepsilon = \eta_5, \qquad (27)$$

we can write Eq. (26) in a similar form similar to Eq. (10). To avoid repetition, we can follow the same procedures as done in the proceeding Eqs. (10) to (19) of Sec. 2.1, we obtain the energy eigenvalues as below

$$E_{nm} = -\frac{\hbar^2}{2a}\left\{\frac{-\frac{eB}{\hbar c}\delta + \frac{\lambda^2 a}{\hbar^2}(V_1+4V_2)}{2\sqrt{\left(\frac{eB}{2\hbar c}\right)^2 + \frac{a\lambda^4}{12\hbar^2}(V_1+16V_2)}} + \frac{1}{8}\left[\frac{a\lambda^3}{3\hbar^2}(V_1+8V_2)\right]^2 - 1 + n\right\}^2 + \frac{\hbar^2}{2a}\delta^2$$
$$+ \frac{3}{8}\frac{\hbar^2}{a} - 2(V_1+V_2) \qquad (28)$$

For the inverse-square mass case, we have showed the energy as a function of the potential parameter $\lambda$ and mass parameter $a$ in Figs. 5 and 6. In fact, in Fig.5, the energy as a function of $\lambda$ is plotted for different values of $B$ and $\Phi$. It is obvious that energy increasing positively with increasing of potential parameter $\lambda$. Further, the particle becomes less positive with increasing $B$ but more repulsive with increasing $\Phi$.

Also in Fig.6, the energy versus $a$ is plotted for different values of $B$ and $\Phi$ with $V_1 = V_2 = 0.01$ and for different values of $V_1$ with $B = \Phi = 2T$. Similar behavior of energy is seen as it decreases with increasing $a$ and particle becomes less positive with increasing $\Phi$. However, it is more strongly repulsive with increasing $B$.

Now let us study the thermodynamic properties of the present model. If we consider the system to be at equilibrium state with a heat bath at a temperature $T$, the canonical partition function is given by

$$Z = \sum_{n,m} e^{\beta \frac{\hbar^2}{2a} \left\{ \frac{-\frac{eB}{\hbar c}\delta + \frac{\lambda^2 a}{\hbar^2}(V_1 + 4V_2)}{2\sqrt{\left(\frac{eB}{2\hbar c}\right)^2 + \frac{a\lambda^4}{12\hbar^2}(V_1 + 16V_2)}} + \frac{1}{8}\left[\frac{a\lambda^3}{3\hbar^2}(V_1 + 8V_2)\right]^2 - 1 + n \right\}^2} \tag{29}$$

$$\times e^{-\beta\left(\frac{\hbar^2}{2a}\delta^2 + \frac{3\hbar^2}{8a} - 2(V_1 + V_2)\right)},$$

where $\beta = 1/k_B T$ with $k_B$ is the Boltzmann constant [46].

At high temperature, we have $\exp\left(-\beta\left(\frac{\hbar^2}{2a}\delta^2 + \frac{3}{8}\frac{\hbar^2}{a} - 2(V_1 + V_2)\right)\right) \approx 1$, then the canonical partition function can be reduced to the form

$$Z = \sum_{n=0}^{n=\zeta} e^{\left[\frac{n-\zeta}{\gamma}\right]^2}, \tag{30}$$

where $\zeta = \dfrac{\frac{eB}{\hbar c}\delta - \frac{\lambda^2 a}{\hbar^2}(V_1 + 4V_2)}{2\sqrt{\left(\frac{eB}{2\hbar c}\right)^2 + \frac{a\lambda^4}{12\hbar^2}(V_1 + 16V_2)}} - \frac{1}{8}\left[\frac{a\lambda^3}{3\hbar^2}(V_1 + 8V_2)\right]^2 + 1$ and $\gamma = \tau/\sqrt{\beta}$ with $\tau = 2a/\hbar^2$ For large $\zeta$ and small $\beta$ at high temperature, we can change the sum in Eq. (30) to an integral as below

$$Z = \gamma \int_0^{\vartheta} e^{t^2} dt = -i\frac{\sqrt{\pi}}{2\sqrt{\beta}}\mathrm{erfi}(\vartheta) = \frac{\sqrt{\pi}}{2\sqrt{\beta}}\mathrm{erf}(\vartheta), \tag{31}$$

where $\vartheta = \zeta\sqrt{\beta}/\tau$, $\mathrm{erf}(\vartheta)$ is the error function for all complex $\vartheta$ and $\mathrm{erfi}(\vartheta) = -i\mathrm{erf}(\vartheta)$ into maple 13.

Thus, from the canonical partition function, we can easily calculate the various thermodynamic variables of the present system like free energy, entropy, specific heat and others as follows:

The internal energy $U$ for the system is obtained

$$U = -\frac{\partial}{\partial \beta} \ln Z \tag{32}$$

$$= \frac{1}{\beta}\left[1 - \frac{\vartheta}{DawsonF(\vartheta)}\right] = -\frac{2\vartheta}{\sqrt{\pi}Erfi(\vartheta)}\left[\frac{e^{\vartheta^2}}{2\beta} - \frac{\sqrt{\pi}Erfi(\vartheta)}{4\beta\vartheta}\right],$$

where Dowson function is defined as $F(\zeta) = e^{-x^2}\int_0^x e^{y^2}dy = \frac{\sqrt{x}}{2}e^{-x^2}erfi(x)$. In Mathematica, the Dawson integral is defined as $DawsonF[\vartheta]$ and the imaginary error function is defined as $Erfi[x]$, and the error function can also be written as $Erf[x] = \frac{2}{\sqrt{\pi}}\int_0^x e^{t^2}dt$ in mathematics [47].

To examine the thermal properties of present model in the inverse-square mass case, we plot internal energy $U$ versus temperature $T$ for different values of $a$ and magnetic field $B$ as shown in Fig. 7 and Fig. 8, respectively. The internal energy increases with increasing temperature, i.e, becomes less attractive. The internal energy becomes less or more attractive when $a$ or $B$ is increasing. In Figure 9, we plot the internal energy $U$ versus $a$ for different values of $B$. The internal energy is less attractive with increasing $a$. However, it becomes more attractive when increasing the strength of magnetic field. Notice that for fixed $a$, the values of the energy decreases when the magnetic field grows.

Similarly one can calculate the specific heat capacity $C_v$ as

$$C_v = k_B\beta^2 \frac{\partial^2}{\partial \beta^2}\ln Z = -k_B\beta^2 \frac{\partial}{\partial \beta}U$$
$$= -\frac{\partial}{\partial \beta}\left\{-\frac{2\vartheta}{\sqrt{\pi}Erfi(\vartheta)}\left[\frac{e^{\vartheta^2}}{2\beta} - \frac{\sqrt{\pi}Erfi(\vartheta)}{4\beta\vartheta}\right]\right\} \qquad (33)$$
$$= \frac{2\vartheta}{\sqrt{\pi}erfi(\vartheta)}\left(-\frac{1}{2}\frac{e^{\vartheta^2}}{\beta^2} + \frac{1}{4}\frac{\sqrt{\pi}erfi(\vartheta)}{\beta^2\vartheta}\right).$$

In Fig. 10, we plot the specific heat capacity $C_v$ versus $a$ for different values of $B$. The specific heat capacity is less positive with increasing $a$. However, increasing magnetic field $B$ makes specific heat more positive.

In Fig. 11, we plot the specific heat capacity $C_v$ versus temperature $T$ for different values of $a$ and $B$. We see that $C_v$ increases exponentially with increasing temperature and becomes more positive with increasing magnetic field but less repulsive with increasing $a$.

The Helmholtz free energy is $= -\ln Z/\beta$. It can be applied to obtain the entropy $S = (U - F)/T$ or $S = k_B\ln Z - k_B\beta\frac{\partial}{\partial \beta}\ln Z$. This yields

$$S = k_B \ln Z - k_B \beta \frac{\partial}{\partial \beta} \ln Z$$

$$= k_B \ln\left(\frac{\sqrt{\pi}}{2\sqrt{\beta}}\text{erf}(\vartheta)\right) - k_B \beta \frac{1}{\sqrt{\pi}.erfi(\vartheta)} 2\left(\frac{\vartheta\, e^{\vartheta^2}}{2\,\beta} - \frac{1}{4}\frac{\sqrt{\pi}}{\beta\sqrt{\beta}}erfi(\vartheta)\right)\sqrt{\beta} = \quad (34)$$

$$\frac{1}{2}k_B\left[1 - \frac{\vartheta}{DowsonF(\zeta)} + 2\log\left(\frac{erfi(\vartheta)}{\sqrt{\beta}}\right) + \log\left(\frac{\pi}{4}\right)\right].$$

In Fig. 12, we show the entropy $S$ decaying exponentially with increasing the mass parameter $a$ for different values of $B$ and $T$. The entropy changes with temperature and magnetic field in similar fashion. It is noticed that entropy decays faster with increasing temperature.

## 3. Discussions

We examine energy behavior with the parameters used in our model. In Fig. 1, we plot ground state energy as a function of mass density parameter $a$ for various values of magnetic field strength, $B = 1.0, 2.0\, Tesla$ when $\Phi = 1.0\, Tesla$. The energy increases exponentially with the increasing of mass distribution parameter $a$. The particle is strongly bound with small values of $a$ while less bound with the increasing of $a$ values. The magnetic field change has a slight effect on energy curve. Further, it is obvious that the energy decreases with increasing $B$ for a fixed value of $a$. In fact, increasing magnetic field makes a particle strongly bound. The effect of the magnetic flux density, $\Phi = 2.0, 3.0\, Tesla$ when $B = 1.0\, Tesla$ on energy is stronger as they become more bound and is very similar, but greater than that of magnetic field. In this case, we also see that the energy is similar of previous case, namely, the energy decreases with increasing $\Phi$ for a fixed value of $a$.

In Fig. 2, the behavior of energy versus $\lambda$ is linear for different values of magnetic field strength $B = 1.0, 3.0,\ 5.0\, Tesla$ and magnetic flus density $\Phi = 1.0, 12.0, 14.0\, Tesla$, respectively.

In Fig. 3 and Fig.4, the energy increases linearly with increasing magnetic field B and magnetic flux density $\Phi$, respectively, for various values of $a = 0, 0.1, 0.2$ The particle becomes less repelling when $a$ is increasing under the influence of both B and $\Phi$. In comparing these Figures, we observe that the effect of the magnetic flux density on the energy is mostly apparent than the effect of the magnetic field.

For the inverse-square mass case, in Fig.5, we plot energy versus $\lambda$ for different values of $B = 2.0, 2.4\,T$ with $\Phi = 2.0T$ and $\Phi = 2.0, 2.4\,T$ with $B = 2.0T$. It is obvious that energy increases positively with increasing of potential parameter $\lambda$ as exponentially. Further, the particle becomes less repulsive with increasing $B$ but more repulsive with increasing $\Phi$.

Also in Fig.6, we plot the energy versus $a$ for different values of $B$ $and$ $\Phi$ with $V_1 = V_2 = 0.01$ and for different values of $V_1$ with $B = 2\Phi = 2T$. Notice that for $a$ fixed value of the energy decreases when the magnetic field grows. But, the energy increases when the magnetic flux and the potential of $V_1$ grow. In fact, similar behavior is noticed, energy decreases with increasing $a$ and particle becomes less repulsive with increasing $\Phi$. However, it is more strongly repulsive with increasing $B$.

On the other hand, to examine the thermal properties of present model in the inverse-square mass case, we plot internal energy $U$ versus temperature $T$ for different values of mass parameter $a = 1.0, 1.2$ and magnetic field $B = 1.0, 10.0\,Tesla$ in Fig. 7 and Fig. 8, respectively. Indeed, when Figs. 7 and 8 is compared together, we observe that effect changing of the mass parameter is more on the internal energy from the magnetic field at temperature $T$. In fact, Also, when the mass parameter increase then the internal energy increase for a fixed of $T$. But, when the magnetic field increases then the internal energy decreases for a fixed value of $T$. Moreover, the internal energy increases with increasing temperature, i.e, becomes less attractive. The internal energy becomes less or more attractive when $a$ or $B$ is increasing. In Figure 9, we plot the internal energy $U$ versus $a$ for different values of $B = 1.0, 10.0\,Tesla$. The internal energy is less attractive with increasing $a$. However, it becomes more attractive when increasing the strength of magnetic field.

In Fig. 10, we plot the specific heat capacity $C_v$ versus $a$ for different values of $B = 1.0, 5.0\,tesla$. The specific heat capacity is less positive with increasing $a$. However, increasing magnetic field $B$ makes specific heat more positive. In Fig. 11, we plot the specific heat capacity $C_v$ versus temperature $T$ for different values of $a$ and $B$. Indeed, in comparing influence two physical quantities such as the mass parameter and magnetic field, we observe that effect changing of the magnetic field is more on the specific heat capacity from the mass parameter at temperature $T$. In fact, Also, when the mass parameter increase then the specific heat capacity decrease for a fixed of $T$. But, when the magnetic field increase then the specific heat capacity increase for a fixed for a fixed of $T$. Moreover, we see that $C_v$ increases exponentially with increasing temperature and becomes more positive with increasing magnetic field but less repulsive with increasing $a$.

In Fig. 12, we show the entropy $S$ decaying exponentially with increasing the mass parameter $a$ for different values of $B$ and $T$. Indeed, in comparing influence two physical quantities such as temperature and magnetic field, we observe that effect changing of the temperature is more on the entropy from the magnetic field. In fact, Also, when the temperature increase then the entropy decrease for a fixed of $a$. But, when the magnetic field increase then the entropy increase for a fixed for a fixed of $a$. However, the entropy changes with temperature and magnetic field in similar fashion. It is noticed that entropy decays faster with increasing temperature.

## 4. Concluding Remarks

We have solved the Schrodinger equation for a particle with a position-dependent mass (PDM) placed in the superposition of Morse and Coulomb potential fields under the influence of external magnetic and AB flux fields. We have calculated the bound state energies and the corresponding wave functions with a suitable change to the dependent variables by using the series method. Our results are reasonable and agree with the results obtained by other authors when mass is stationary and found to be highly in good agreement. Our results of the energy states are plotted in Figures 1 to 6. In comparing these Figures, we observe that the effect of the magnetic flux density on the energy is more apparent than that of the magnetic field.

We see the great effect of potential parameter $\lambda$ and mass density parameter $a$ on the energy. On the other hand, we studied the thermodynamic properties of the potential model with inverse-square mass case. The internal energy $U$ becomes less or more attractive when $a$ or $B$ is increasing. The internal energy is less attractive with increasing $a$. However, it is more attractive when increasing the strength of magnetic field.

Indeed, in comparing the influence of two physical quantities such as temperature and magnetic field, we observe that effect of changing the temperature is more on the internal energy, specific heat and entropy from the magnetic field. Also, as temperature increases then the entropy decreasing for a fixed value of $a$. But, when the magnetic field increase then the entropy increase for a fixed for a fixed of $a$.

## Conflict of Interests

The author(s) declare(s) that there is no conflict of interest regarding the publication of this paper.

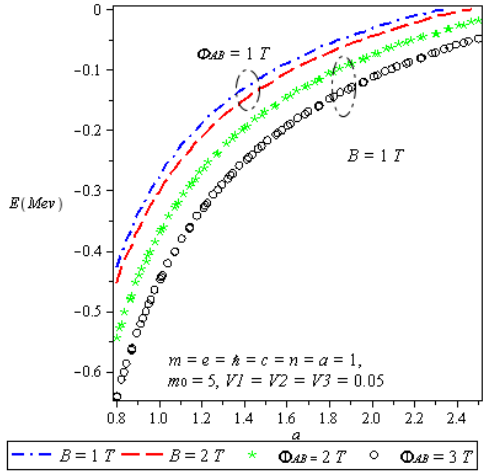

**Figure 1** The energy versus $a$ for different values of magnetic field stregth $B$ and magnetic flux density $\Phi_{AB}$.

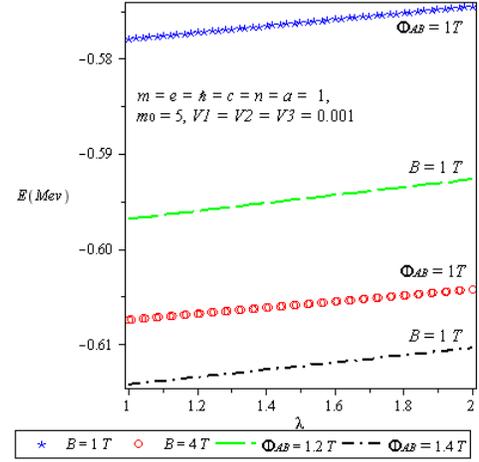

**Figure 2** The energy versus $\lambda$ for different values of magnetic field stregth $B$ and magnetic flux density $\Phi_{AB}$

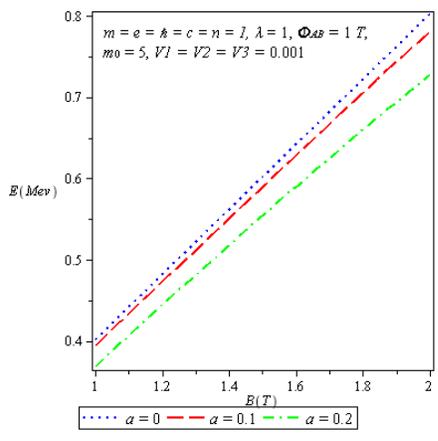

**Figure 3** The energy versus $B$ for different values of mass density $a$

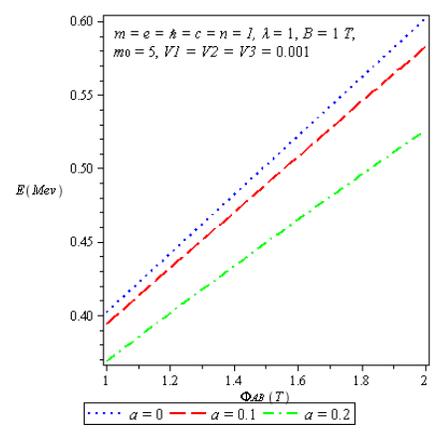

**Figure 4** The energy versus $\Phi$ for different values of mass density $a$

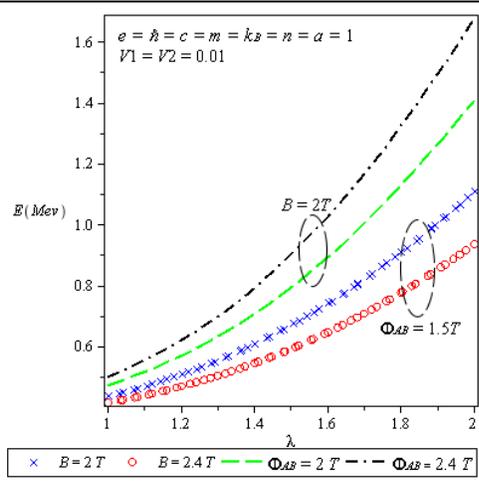

**Figure 5** The energy versus $\lambda$ in different values of $B$ and $\Phi_{AB}$ in the inverse-square mass case.

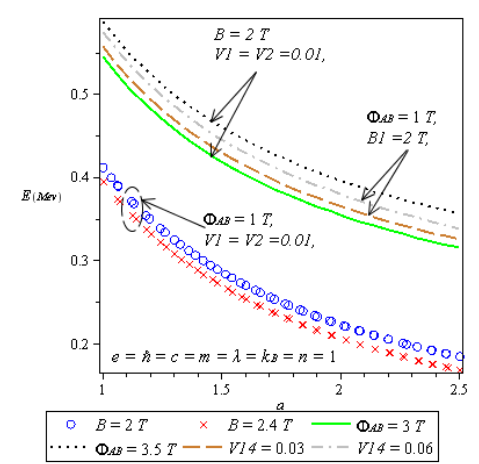

**Figure 6** The energy versus $a$ for different values of $\Phi_{AB}$ and $V_1$ in the inverse-square mass case.

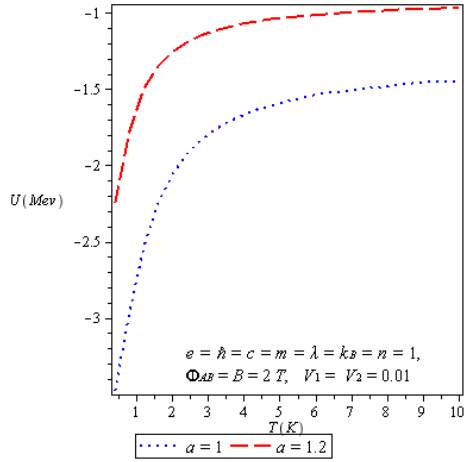

**Figure 7** The internal energy $U$ versus $T$ for different values of $a$ for second case,

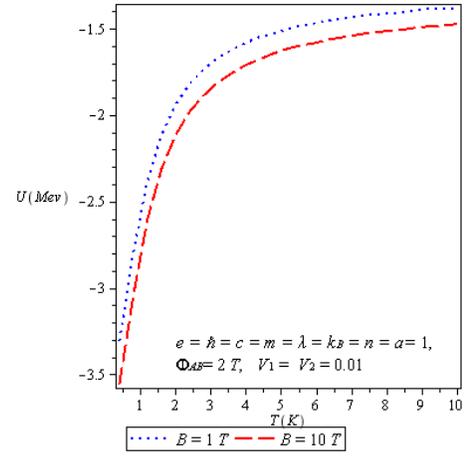

**Figure 8** The internal energy $U$ versus $T$ for different values of $B$ for second case.

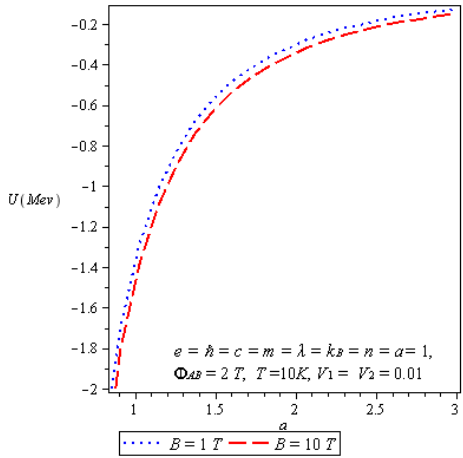

**Figure 9** The internal energy $U$ versus $a$ for deferent value of $B$ for second case.

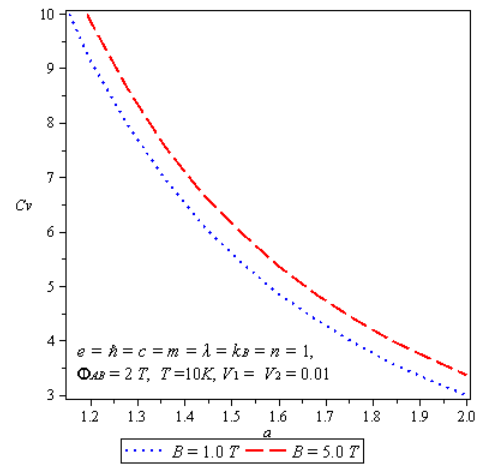

**Figure 10** The specific heat capacity $C_v$ versus $a$ for different values of $B$ for second case.

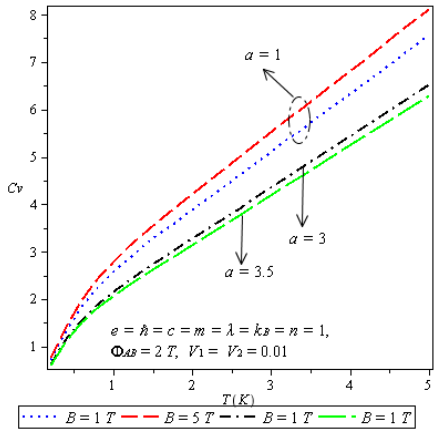

**Figure 11** The specific heat capacity $C_v$ versus $T$ for different values of $a$ and $B$ for second case.

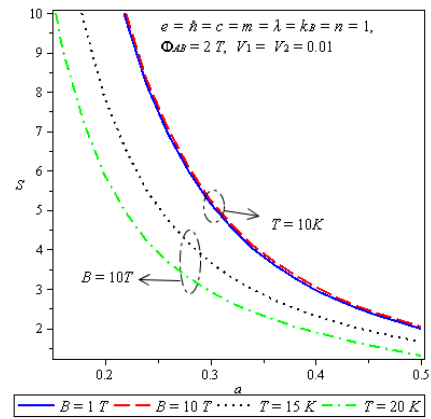

**Figure 12** The entropy $S$ versus $a$ for different values of $B$ and $T$ for second case.